\documentclass[prl,reprint,twocolumn,showpacs,preprintnumbers,amsmath,amssymb,nofootinbib,tightenlines,groupedaddress,superscriptaddress]{revtex4-1} 
\usepackage{graphicx}

\usepackage{mathrsfs}
\usepackage{hyperref}

\usepackage{color}
 \usepackage[utf8]{inputenc}

\begin{document}

\title{Efficient integrand reduction for particles with spin}

\author{Rutger H.  Boels}
\email{Rutger.Boels@desy.de}
\affiliation{II. Institut f\"ur Theoretische Physik, Universit\"at Hamburg,  \\ Luruper Chaussee 149, D-22761 Hamburg, Germany}

\author{Qingjun Jin}
\affiliation{CAS Key Laboratory of Theoretical Physics, Institute of Theoretical Physics, \\ Chinese Academy of Sciences, Beijing 100190, China}

\author{Hui Luo}
\affiliation{II. Institut f\"ur Theoretische Physik, Universit\"at Hamburg,  \\ Luruper Chaussee 149, D-22761 Hamburg, Germany}
\affiliation{PRISMA Cluster of Excellence, Johannes Gutenberg University, \\ Staudingerweg 9, 55128 Mainz, Germany}

\date{\today}

\preprint{MITP/18-013}

\begin{abstract}
Scattering amplitudes with spinning particles are shown to decompose into multiple copies of simple building blocks to all loop orders, which can be used to efficiently reduce these amplitudes to sums over scalar integrals. Absence of unphysical kinematic singularities cleanly exposed by the method uncover novel consistency relations among master integrals and their coefficients. Analytic results are obtained for the five gluon, two loop, and four gluon, three loop planar scattering amplitudes in pure Yang-Mills theory as well as for leading singularities to even higher orders.
\end{abstract}

%12.38.Bx 	Perturbative calculations (QCD category)



\maketitle

\section{Introduction}
Precise predictions for experiments are the backbone of physics. In particle physics these take the form of scattering cross-sections, which are assembled out of scattering amplitudes that are computed perturbatively, as well as experimentally determined input such as parton distribution functions. This inherent interest in computing scattering amplitudes is becoming ever more important due to the absence of a smoking-gun observation of physics  beyond the well-established standard model at the Large Hadron Collider (LHC) at CERN: since the energy frontier will not move in the short term, precision physics is the most probable vector for near future discovery. The frontiers of the state-of-the-art are loosely measured in the number of external particles and internal loops. For non-supersymmetric Yang-Mills theory, very recently first (semi-)numerical computations of the planar five gluon amplitudes at two loops were reported in \cite{Badger:2017jhb, Abreu:2017hqn}, while analytically a special, equal helicity amplitude is known at two loops through seven external gluons \cite{Badger:2013gxa, Badger:2016ozq, Dunbar:2017nfy}.

Beyond phenomenological interest, scattering amplitudes also attract formal interest as a basic output of quantum field or string theory which displays structures which may not be obvious from their original formulation. A prime example of this are the Kawai-Lewellen-Tye \cite{Kawai:1985xq} relations discovered first in string theory, which relate a certain sum over products of gluon scattering amplitudes to graviton scattering amplitudes. These are referred to generally as ``double copy'' type relations, see e.g. \cite{Bern:2008qj}. Scattering amplitudes obey in general physical constraints such as gauge and global symmetries, locality and unitarity, which are to an extent mutually redundant \cite{Arkani-Hamed:2016rak} \cite{Rodina:2016mbk}. In \cite{Boels:2016xhc}, drawing on ideas in \cite{Barreiro:2013dpa}, it was shown these constraints can be solved systematically. In \cite{Boels:2017gyc} (see also \cite{Bern:2017tuc}) specific solutions for four particle scattering were constructed and used to (re-)compute loop level scattering amplitudes with gluons and gravitons using unitarity. Earlier Feynman-graph based computations in \cite{Glover:2003cm, Gehrmann:2011aa} as well as the well-known computation of the gyromagnetic factor \cite{Peskin:1995ev} employ similar technology. Algebraic complexity has prohibited practical applicability thus far. Here this is solved by obtaining solutions to physical constraints as certain multi-copies of simpler one and two gluon building blocks. 

The resulting coefficients and integrals are typically reduced further by using the linear integration-by-parts (IBP) identities to express the amplitudes in a much smaller basis of so-called master integrals. This step is a very well-known bottleneck due to its overwhelming intermediate complexity. The multi-copy basis allows a particularly clean view on this reduction and its output. As a result, we uncover the need for intricate relations among the coefficients of the different masters which follow from the absence of residues at non-physical poles. These relations can be derived from differential equations and can be used as powerful internal consistency check as well as as a tool to derive integral coefficients.  

To demonstrate the potential of the methods explored here we showcase analytic applications to the planar five gluon, two loop amplitude and the planar four gluon, three loop amplitudes in pure Yang-Mills theory, expressed in a master integral basis; the five point result is closely related to the very recent semi-numerical results in \cite{Badger:2017jhb, Abreu:2017hqn}. Furthermore, we briefly show how to extend our techniques to massive matter. Throughout this Letter we work in dimensional regularisation to regulate the divergent loop integrals, in the scheme where all internal and external particles are in $d$ dimensions.

\section{External kinematics from a multi-copy}
Scattering amplitudes with spinning matter are Lorentz scalars and little group tensors. Every particle is associated with a polarisation tensor, which embeds a copy of the appropriate little group representation into the Lorentz group. For massless bosons for instance, which will constitute our main example for illustration, this involves products of the polarisation vector $\xi_{\mu}^I(p)$ where the Roman index indicates a little group and the greek index a Lorentz vector respectively. The polarisation vectors have to obey transversality, $p^{\mu} \xi_{\mu}^I(p) = 0$, and the corresponding scattering amplitude on-shell gauge invariance \cite{Noether:1918zz},
\begin{equation}
 A(\{\xi_i\rightarrow p_i\}) \rightarrow 0\, ,
\end{equation}
for each individual massless particle of the amplitude. In addition, there is momentum conservation for the external momenta, which are all taken here to be complex and inward pointing. Since Poincare symmetry is exact, scattering amplitudes are multilinear to all orders in the polarisation vectors. All the mentioned constraints are therefore linear and can be solved at least in principle, as explained in \cite{Boels:2016xhc}. The solution space is spanned by a set of tensor structures. Every scattering amplitude can be expressed as a linear combination of these solutions,
\begin{equation}
\mathcal A = \sum_{i} \alpha_i B_i \, .
\end{equation}
In this form the scalar coefficients involve only (integrals over) Lorentz invariants constructed out of internal and external momenta. Given any form of $A$, the coefficients can be determined from multiplication with $B_j$ and summing over all helicities,
\begin{equation}\label{eq:Pmatrixfirst}
 \sum_{\textrm{helicities}}  B_j \mathcal A = \sum_{i} \alpha_i \left( \sum_{\textrm{helicities}}  B_j B_i \right) \equiv \sum_i P_{ji} \alpha_i \, ,
\end{equation}
which gives a scalar, linear problem after using the completeness relation to sum over helicities,
\begin{equation}
\sum_{\textrm{helicities}} \xi_{\mu} \xi_{\nu} = \eta_{\mu\nu} - \left(\frac{p_{\mu} q_{\nu} + p_{\nu} q_{\mu} }{ q\cdot p} \right)\, ,
\end{equation}
where $q$ is a gauge choice that drops out of the result. The matrix $P_{ji}$ is invertible in general as the $B$ form a basis. However, as can be seen from the table in \cite{Boels:2016xhc}, the size of the spaces involved grows very quickly: six gluon scattering already involves $2364$ solutions! Computing, let alone inverting, a polynomial matrix of this size is generally unfeasible. 

This problem is solved here by a good basis choice. Consider the parity even scattering of one gluon and $n-1$ scalar particles. There are $n-2$ independent contractions of external momenta with the polarisation vector of the single gluon. Gauge invariance yields one constraint. The solution space of dimension $n-3$ can be spanned by objects 
\begin{equation}
A_i(j,k) =  (p_k \cdot p_i) \,p_j \cdot \xi_i  -  (p_j \cdot p_i) \,p_k \cdot \xi_i \, ,
\end{equation}
for instance by the set
\begin{equation}\label{eq:Aset}
\{ A_i(j) = A_i(i+j, i+j+1) | j \in \{1,\ldots, n-3 \} \} \, ,
\end{equation}
with particle momenta identified cyclically. Next, consider the scattering of two gluons and $n-2$ scalar particles. A special class of solutions to the physical constraints is given by multiplying copies of the solutions found in the single gluon case:
\begin{equation}
A_1(j) A_2(k) \qquad j,k \in \{1,\ldots, n-3 \} \, .
\end{equation}
This set is however not complete, as there is also
\begin{equation}
 C_{i,j} = (\xi_{i} \cdot \xi_j)(p_i \cdot p_j)- (p_i \cdot \xi_j) (p_j \cdot \xi_i) \, ,
 \end{equation}
which is proportional to two contracted linearised field strength tensors, $F_{\mu\nu}(\xi_1) F^{\mu\nu}(\xi_2)$. The inner product of polarisation vectors makes it manifestly independent from the set of  two copies of $A$'s. No additional solutions exist in the two gluon case. For more gluons, a set of solutions can always be obtained by multiple copies of lower gluon number solutions. We conjecture this set constructed of all possible $A$ and $C$ type building blocks for a given number of gluons is both linearly independent and complete in general dimensions. This was explicitly checked through six gluon amplitudes. The total number of basis elements with $n$ gluons and no scalars is
\begin{equation}
N_n = \sum_{k=0}^{\left \lfloor{n/2}\right \rfloor } \frac{n! (n-2)^{(n-2 k)}}{2^k k! (n-2 k)!} \, ,
\end{equation}
which agrees with the numbers obtained in \cite{Boels:2016xhc} through $n=7$. 

\begin{figure}[t]
\includegraphics[scale=0.3]{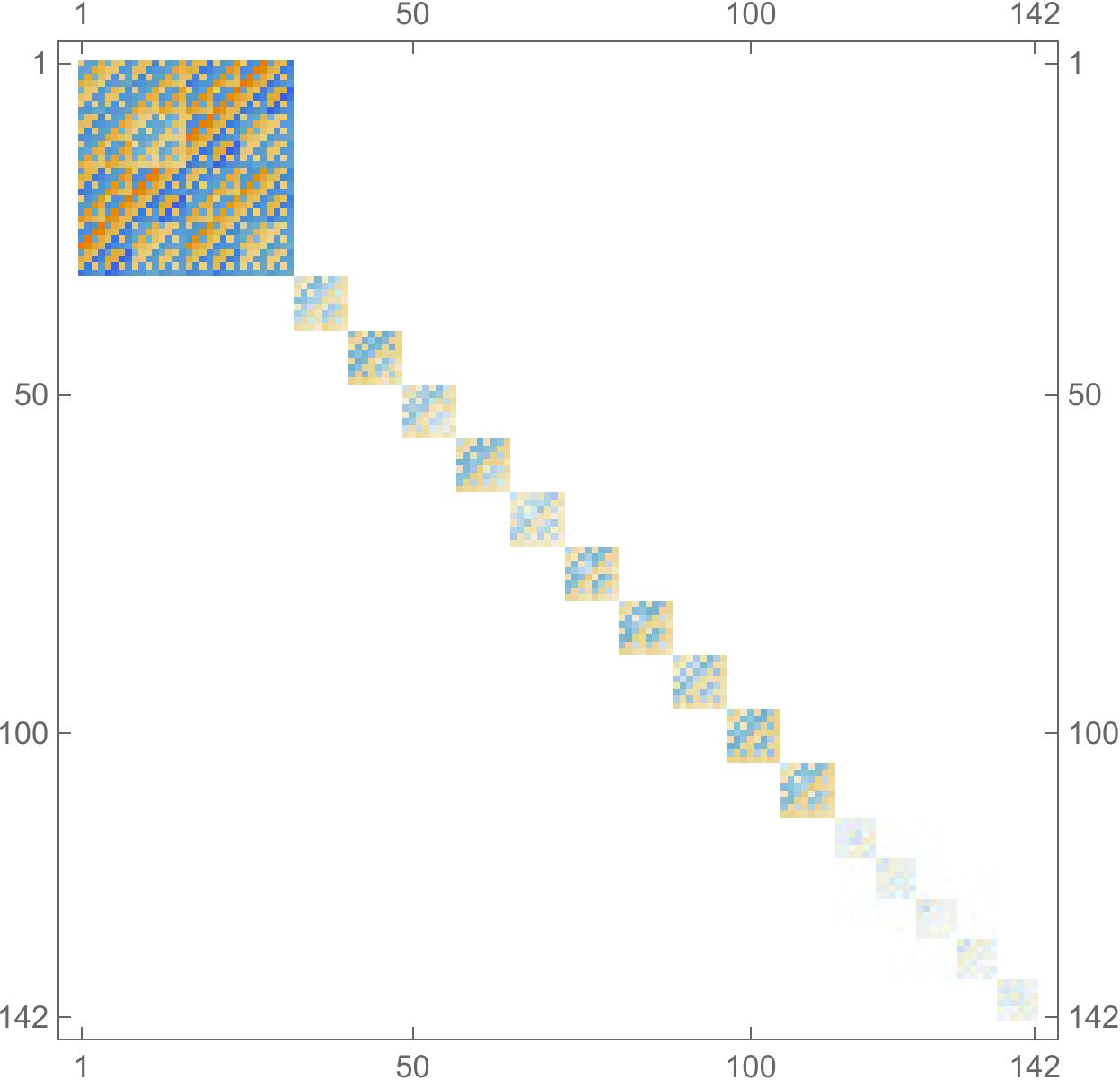}
\caption{\label{fig:5ptPmatrillust}P-matrix for five gluons for the basis constructed from $A$ and $D$ tensors, evaluated on random integer values.}
\end{figure}

To solve equation \eqref{eq:Pmatrixfirst} first construct a new tensor 
\begin{equation}
D_{i,j} = C_{i,j} - \sum_{k,l=1}^{n-3} X_{ij}(k,l) A_{i}(k) A_{j}(l) \, ,
\end{equation}
and require that it is ``orthogonal'' to the A-tensors as
\begin{equation}
\sum_{h_i} A_{i}(k) D_{i,j} = 0 =  \sum_{h_j} A_{j}(k) D_{i,j},\,\, \forall k,   \, ,
\end{equation}
by summing over the helicities of particles $i$ or $j$ respectively. To find the unique solution for $X$, first consider 
\begin{equation}
P^A_i(k,l) = \sum_{h_i} A_{i}(k) A_{i}(l) \, .
\end{equation}
The set $A$ in equation \eqref{eq:Aset} is mapped to linear combinations of itself by Bose-permutations of the other legs than $i$. This severely constrains the matrix $P^A_i$, as well as its inverse. Now construct the dual vector or projector
\begin{equation}
A^i(k) \equiv \sum_l (P^A_i)^{-1}(k,l) A_i(l) \, ,
\end{equation}
which obeys
\begin{equation}
A^i(k) A_i(l) \equiv \sum_{\textrm{helicities},i } A^i(k) A_i(l) = \delta(k,l) \, .
\end{equation}
With this notation,
\begin{equation}
D_{i,j} = C_{i,j} - \sum_{k,l=1}^{n-3} A_i(k) A_j(l) \left(A^m(k) A^n(l) C_{m,n}\right) \, , 
\end{equation}
holds. For a given subset of legs $s$ with $|s|$ elements ($|s|$ is even) one can construct all $(|s|-1)!!$ matrices obtained by multiplying $\frac{|s|}{2}$ $D$ matrices, which we will denote $D^{s}$. For this set consider
\begin{equation}
P^{D^{s}}_{ij} = \sum_{\textrm{helicities}} D^{s}_i  D^{s}_j,  \quad i,j=1,\dots, (|s|-1)!! \, .
\end{equation}
This computation is straightforward as for instance
\begin{equation}
\sum_{\textrm{helicities}}D_{i,j} D_{i,j} = (p_i \cdot p_j)^2 (d-n+1) \, , %\qquad D_{i,j} D^{j,k} = \frac{(p_i \cdot p_j) (p_k \cdot p_j) }{p_i \cdot p_k} D_{il} \delta^{lk} 
\end{equation}
holds. Bose permutations simply act on the labels of the $D$ matrices and therefore permute the entries of the $P^{D^{s}}$-matrix as well as its inverse: a small number of elements determines the whole matrix. The inverse of this factor is observed up to at least $6$ particles to be very simple: the entries of the inverse are the entries of the original matrix inverted up to a function of $d$. This allows one to efficiently construct a projector, $D^{s,i}$, such that
\begin{equation}
\sum_{\textrm{helicities}} D^{s,i} D^{s}_j = \delta^i{}_j \, .
\end{equation}
Since the $D$ and $A$ tensors are orthogonal, the projectors onto specific basis elements factorise into sets with different numbers of $D$ tensors as well as within these sets into choices of which particles are on the $A$-type tensors, see figure \ref{fig:5ptPmatrillust} for an illustration in the five point case. Individual projectors can efficiently be computed, which was explicitly checked through six points.

%%%%%%%%%%%%%%%%%%%%%%%%%%%%%%%%%%%%%%%%%%%%%%%%%%%%%%%%%%%%%%%
%%%%%%%%%%%%%%%%%%%%%%%%%%%%%%%%%%%%%%%%%%%%%%%%%%%%%%%%%%%%%%%

\section{Necessity of relations among master integral coefficients}
After using the projectors on a given loop amplitude one obtains a sum over scalar integrals,
\begin{equation}
\left(\mathcal I\right)_{\textrm{proj}} = \sum_k \int  d\, l_i \frac{f_k(l_i, p_j)}{\prod K_m} \, ,
\end{equation}
%%%
where $f$ is a function of all independent inner products of external and internal momenta. The propagators $K_i$ are scalar functions, quadratic in internal momenta. Coordinates to identify integrals have to be chosen, but are only determined up to linear shifts of the loop momenta. Infinitesimally, this freedom induces so-called integration by parts (IBP) identities \cite{Chetyrkin:1981qh} among the integrals. The linear IBP identities can be solved systematically by Gaussian elimination after choosing an ordering on the vector space \cite{Laporta:2001dd}, basically aiming to solve complicated integrals in terms of simpler ones. Several public codes exist to perform this step such as {\tt FIRE} \cite{Smirnov:2008iw, Smirnov:2013dia, Smirnov:2014hma}, KIRA \cite{Maierhoefer:2017hyi}, {\tt Reduze} \cite{vonManteuffel:2012np} and {\tt LiteRed} \cite{Lee:2012cn, Lee:2013mka}. The output is a sum over a much smaller basis of integrals referred to as ``master integrals"
\begin{equation}
\left(\mathcal I\right)_{\textrm{proj}} = \sum_i c_i {\rm{MI}}_i \, .
\end{equation}
As emphasised in \cite{Boels:2017gyc} (see also \cite{BjerrumBohr:2007vu}), the poles of these coefficients can contain unphysical poles whose residue has to vanish between contributions of different master integrals. This type of singularity is known to occur in differential equations for the master integrals w.r.t. a Mandelstam invariant, 
\begin{equation}
\frac{\partial {\rm{MI}}_i}{\partial s} = M_i{}^j \, {\rm{MI}}_j \ ,
\end{equation}
see for instance \cite{Henn:2014qga}. The matrix $M$ is a function of the external kinematic invariants and the dimension. If all masters are finite in a certain kinematic limit (or already matched on all logarithmic singularities), say $u\rightarrow 0$, the differential equation can be used to derive a series of relations from a Laurent expansion.

As an example, consider the case of a massive one loop contribution of a scalar or fermion to the color-ordered four gluon amplitude with ordering $1234$. There are $6$ master integrals. The physical problem does not allow poles in the cross-channel pole, $u=(p_1+p_3)^2$. The coefficients of the master integrals except the massive tadpole integral can be determined from unitarity cuts and their explicit form contains non-physical poles up to $\frac{1}{u^4}$. The massive tadpole coefficient cannot be determined from unitarity cuts, see e.g. \cite{Badger:2017gta}. Using the differential equation-derived constraints on the Laurent expansion around $u=0$ shows that the residues of the unphysical poles cancel between the unitarity derived master integrals down to $u^0$, \emph{up to the undetermined massive tadpole coefficient}. Taken together with gauge invariance this fixes the massive tadpole coefficient, up to a freedom associated with coupling constant renormalisation. In this very simple example, one can also study the large mass limit. For more complicated cases it is likely advantageous to transform the differential equations to so-called Fuchsian form (see e.g. \cite{Henn:2013pwa}) where all kinematic singularities are simple poles, for instance using \cite{Gituliar:2017vzm}.

\section{Application to gluon scattering amplitudes}
\begin{figure}[t]
\includegraphics[scale=0.25]{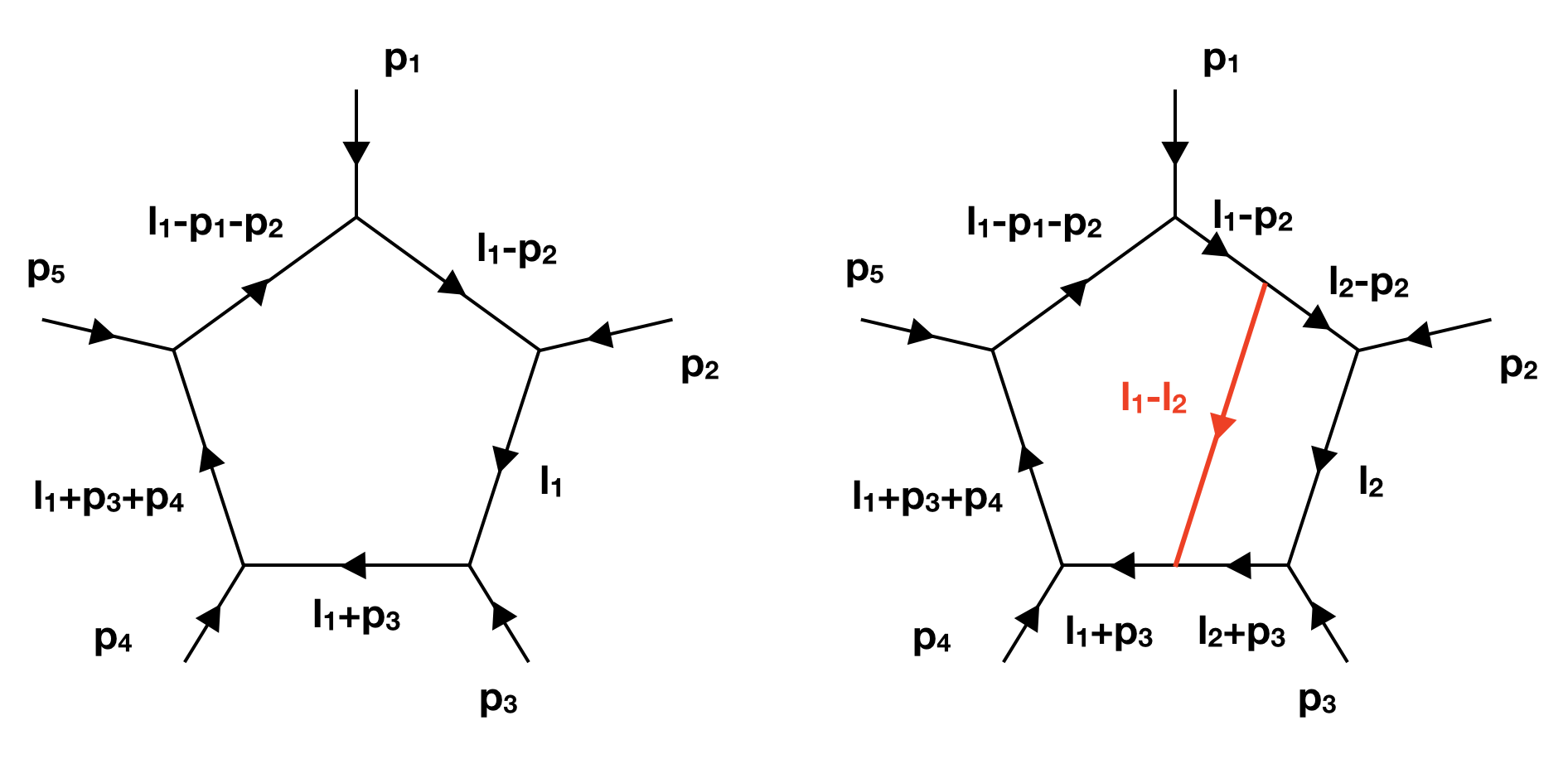}
\caption{\label{fig:5ptillus} Deriving a two loop integral parametrisation of a five point scattering problem from a one loop topology.}
\end{figure}

A good coordinate system to parametrise loop integrals is crucial. For planar integrals integrals with $l$ loops and $n$ legs choose $l$-copies of``n-gon''  one loop integrals, adding the squares of differences of loop momenta, $(l_i - l_j)^2$ for $i<j$. This choice minimises the number of different internal momenta per propagator. In the one-loop basis one should minimise the number of external momenta per propagator. Integral labels can be derived by adding internal propagators to the one-loop progenitor, see figure \ref{fig:5ptillus}. Extensions to massive  matter for planar amplitudes are straightforward. 

%%%%%%%%%%%%%%%%%%%%%%%%%%%%%%%%%%%%%%%%%%%%%%%%%%%%%%%%%%%%%%%
%%%%%%%%%%%%%%%%%%%%%%%%%%%%%%%%%%%%%%%%%%%%%%%%%%%%%%%%%%%%%%%

\subsection{Planar, five point, two loops}
A good set of integral coordinates  is
\begin{multline}\label{eq:fivepointtwoloop}
\left\{(l_1 - l_2)^2, (l_1)^2, (l_1 - p_2)^2,( l_1 - p_1 - p_2)^2, \right. \\
( l_1 + p_3 + p_4)^2, ( l_1 + p_3)^2, (l_2)^2, (l_2 - p_2)^2,\\
\left.( l_2 - p_1 - p_2)^2,( l_2 + p_3 + p_4)^2, ( l_2 + p_3)^2 \right\} \, .
\end{multline} 
The separate pentagons are chosen to never have more than two external momenta. This set has a single manifest graph symmetry since the map $l_1 \leftrightarrow l_2$ induces a permutation of the elements. A cyclic transformation $p_i \rightarrow p_{i+1}$ as well as an inversion such as $(p_1,p_2,p_3,p_4,p_5) \rightarrow (p_5,p_4,p_3,p_2,p_1)$ induce a permutation of the elements after a shift of the loop momentum. The latter transformations do not leave integrals invariant, but map between different integrals parameterised with the above coordinates. 

Proceed by a unitarity computation as outlined in \cite{Boels:2017gyc} taking inspiration from especially \cite{Bern:1994cg}. This requires computation of all cyclically independent cuts by multiplying tree amplitudes, summing over internal polarisations and projecting on the external kinematics using the basis constructed above. The needed tensor algebra is straightforward to implement even on a laptop.  There are $142$ tensor basis coefficients. The IBP identities can be solved using FIRE, running on large computing resources (details on request). The most complicated reductions needed are ``pentabox" integrals with five powers of irreducible numerators. Although all five distinct pentaboxes can be parametrised by the coordinates in equation \eqref{eq:fivepointtwoloop}, one of these contains less terms in its propagators than the other four, see the figure \ref{fig:5ptillus}. This seemingly minor simplification was for us key to solve the IBP identities. The master integrals for the case at hand are expressible in harmonic polylogarithms, \cite{Gehrmann:2015bfy} \cite{Papadopoulos:2015jft}.

Possible cross-channel poles of the coefficients can be read off from the projectors for the basis choice,
\begin{equation}\nonumber
A_i \propto \frac{1}{G (p_i + p_{i+1})^2 (p_i + p_{i-1})^2} \quad D_{i,j} \propto  \frac{ 1}{G (p_i + p_{j})^2} \, ,
\end{equation}
where $G$ is the determinant of the Grammian matrix for the independent momenta. 

Cyclic symmetry can be used to simplify the unitarity computation. For the choice of tensor structure basis above,  cyclicity and inversion act as a permutation on the basis. A complete set of master integrals with the same property for cyclic permutations can be chosen, forming a set with $31$ orbits of length $5$. If the coefficient of one representative of a particular orbit is known for all tensor structures, then the coefficients of the other integrals in this orbit can be obtained from cyclic symmetry. There are four topologically distinct cuts: one triple cut and three double double cuts. The triple cut determines $27$ of the orbits, while the double cuts then can be used to fix the remaining four orbits. The thus-obtained results have been verified to match to those of \cite{Abreu:2017hqn, Badger:2017jhb} on a phase-space point by reproducing the master integral coefficients  of the result in \cite{Abreu:2017hqn} for all independent helicity configurations.

\subsection{Planar, four point, three loops}
A good set of integral coordinates  is
\begin{multline}
\left\{( l_1 - l_2)^2,( l_2 - l_3)^2,( l_1 - l_3)^2,\right.  \\
( l_1)^2, ( l_1 - p_3)^2, ( l_1 + p_1 + p_2)^2, ( l_1 + p_2)^2,\\
( l_2)^2, ( l_2 - p_3)^2, ( l_2 + p_1 + p_2)^2, ( l_2 + p_2)^2,\\
\left. (l_3)^2, ( l_3 - p_3)^2, ( l_3 + p_1 + p_2)^2, ( l_3 + p_2)^2 \right\} \, .
\end{multline} 
This set has a manifest order $6$ permutation symmetry from exchanging the three loop momenta. A cyclic transformation $p_i \rightarrow p_{i+1}$ or an inversion $(p_4 \leftrightarrow p_1), (p_2 \leftrightarrow p_3)$ induces a permutation of the elements of this basis for integrals which follows directly from their one loop origin.

The factorised external kinematics project (cuts of) loop amplitudes onto scalar basis coefficients. There are for four gluons $10$ different basis coefficients, falling into $5$ distinct orbits of the cyclic group. The projectors are the source of cross-channel poles, where both $A$ and $D$ are proportional to the determinant of the Grammian matrix for all independent external momenta which here is the product of Mandelstam invariants, $\propto \, s t u $. FIRE5 can be used to address the IBP problem. There are in total $81$ master integrals, for which an analytic expression is known in principle \cite{Henn:2013fah}.

\subsection{Planar and non-planar leading singularities through four loops}
An interesting quantity that has played a central role in maximally supersymmetric Yang-Mills theory such as the amplituhedron \cite{Arkani-Hamed:2013jha} picture has been the notion of leading singularity \cite{Cachazo:2008vp}: the residue of a scattering amplitude after cutting all propagators. This is for pure Yang-Mills theory as well as for Einstein-Hilbert gravity a product of three point amplitudes, sewn together according to a cut trivalent graph at the appropriate loop order. Since cutting external propagators corresponds to tree singularities, one can limit to cutting $1$-particle irreducible graphs. We have checked, using the graph generator DiaGen \cite{diagen}, that the tensor algebra for all (planar and non-planar) leading singularities of the five point three loop and four point four loop gluon amplitudes is straightforward. This shows tensor algebra is no longer a bottleneck for any further progress.

\section{Discussion}
This Letter presents fresh insight into an old problem: how to compute scattering amplitudes in physically interesting theories. In particular, the factorised multi-copy basis constructed here offers cleanly exposes kinematic singularities in master integral coefficients which clearly deserve further study. Most pressing is the question to what extend leading singularities determine complete amplitudes. By the relations uncovered in this Letter, these coefficients certainly constrain the non-leading-singularity master integrals. Since the relations require only knowledge of differential equations, this involves a much simpler IBP reduction problem, potentially bypassing a current bottleneck for computation. Knowing in advance the kinematic singularity structure of master integral coefficients alone is useful. It would furthermore be interesting to explore the connection through Fuchsian form differential equations to the activity launched by \cite{Henn:2013pwa}: currently much more is known about integrals than about complete scattering amplitudes. 

Although the examples in this Letter have focused on planar amplitudes of massless gluons, extensions to more general matter are straightforward. Massive scalars are already covered by the analysis above. An easy further conjecture is that graviton scattering amplitudes can be expressed in a factorised tensor structure basis as well, including $D$-type elements combining the `left' and `right' polarisations. This is certainly correct for four point amplitudes, with $513$ basis elements and a factorised inverse of the P-matrix constructed as for gluons. Massive matter, a subject scratched at the surface here for tadpole coefficients, should be the focus of major attention. Experimental motivation also motivates the study of complete cross-sections: the presented technology can elucidate analytic computations, especially for the needed intricate cancellation of divergences beyond the one loop order. The structure of the integral coordinates is intriguing in this context.  

While this Letter already presents cutting edge applications for analytic computations, there is considerable room for improvement especially for IBP reduction. Our results were obtained with an older public code, paired with our observation on good integral coordinates for IBP reduction. Several groups have been working on IBP reductions with promising first results especially where finite-field methods and cuts are employed, see e.g. \cite{Ita:2015tya, Larsen:2015ped, Georgoudis:2016wff, Abreu:2017hqn, Boehm:2017wjc, Boehm:2018fpv}. Combining our insights for tensor structures with these developments has the potential to truly revolutionise calculational power for explicit as well as collider-relevant quantum field theory predictions.

\begin{acknowledgments}
\section*{Acknowledgements}
The authors would like to thank Yang Zhang for discussions as well as Ben Page and Harald Ita for help in numerical confirmation of results for planar five point two loop amplitudes. RB would like to thank Vladimir Smirnov, Sven-Olaf Moch, Oleksandr Gituliar and Bernd Kniehl for feedback and encouragement. QJ is supported in part by the Chinese Academy of Sciences (CAS) Hundred-Talent Program, by the Key Research Program of Frontier Sciences, CAS, and by Project 11647601 supported by NSFC. This work was supported by the German Science Foundation (DFG) within the Collaborative Research Center 676 ``Particles, Strings and the Early Universe". 
\end{acknowledgments}

\bibliography{bib.bib}

%merlin.mbs apsrev4-1.bst 2010-07-25 4.21a (PWD, AO, DPC) hacked
%Control: key (0)
%Control: author (72) initials jnrlst
%Control: editor formatted (1) identically to author
%Control: production of article title (-1) disabled
%Control: page (0) single
%Control: year (1) truncated
%Control: production of eprint (0) enabled
\begin{thebibliography}{43}%
\makeatletter
\providecommand \@ifxundefined [1]{%
 \@ifx{#1\undefined}
}%
\providecommand \@ifnum [1]{%
 \ifnum #1\expandafter \@firstoftwo
 \else \expandafter \@secondoftwo
 \fi
}%
\providecommand \@ifx [1]{%
 \ifx #1\expandafter \@firstoftwo
 \else \expandafter \@secondoftwo
 \fi
}%
\providecommand \natexlab [1]{#1}%
\providecommand \enquote  [1]{``#1''}%
\providecommand \bibnamefont  [1]{#1}%
\providecommand \bibfnamefont [1]{#1}%
\providecommand \citenamefont [1]{#1}%
\providecommand \href@noop [0]{\@secondoftwo}%
\providecommand \href [0]{\begingroup \@sanitize@url \@href}%
\providecommand \@href[1]{\@@startlink{#1}\@@href}%
\providecommand \@@href[1]{\endgroup#1\@@endlink}%
\providecommand \@sanitize@url [0]{\catcode `\\12\catcode `\$12\catcode
  `\&12\catcode `\#12\catcode `\^12\catcode `\_12\catcode `\%12\relax}%
\providecommand \@@startlink[1]{}%
\providecommand \@@endlink[0]{}%
\providecommand \url  [0]{\begingroup\@sanitize@url \@url }%
\providecommand \@url [1]{\endgroup\@href {#1}{\urlprefix }}%
\providecommand \urlprefix  [0]{URL }%
\providecommand \Eprint [0]{\href }%
\providecommand \doibase [0]{http://dx.doi.org/}%
\providecommand \selectlanguage [0]{\@gobble}%
\providecommand \bibinfo  [0]{\@secondoftwo}%
\providecommand \bibfield  [0]{\@secondoftwo}%
\providecommand \translation [1]{[#1]}%
\providecommand \BibitemOpen [0]{}%
\providecommand \bibitemStop [0]{}%
\providecommand \bibitemNoStop [0]{.\EOS\space}%
\providecommand \EOS [0]{\spacefactor3000\relax}%
\providecommand \BibitemShut  [1]{\csname bibitem#1\endcsname}%
\let\auto@bib@innerbib\@empty
%</preamble>
\bibitem [{\citenamefont {Badger}\ \emph
  {et~al.}(2017{\natexlab{a}})\citenamefont {Badger}, \citenamefont
  {Br{\o}nnum-Hansen}, \citenamefont {Hartanto},\ and\ \citenamefont
  {Peraro}}]{Badger:2017jhb}%
  \BibitemOpen
  \bibfield  {author} {\bibinfo {author} {\bibfnamefont {S.}~\bibnamefont
  {Badger}}, \bibinfo {author} {\bibfnamefont {C.}~\bibnamefont
  {Br{\o}nnum-Hansen}}, \bibinfo {author} {\bibfnamefont {H.~B.}\ \bibnamefont
  {Hartanto}}, \ and\ \bibinfo {author} {\bibfnamefont {T.}~\bibnamefont
  {Peraro}},\ }\href@noop {} {\  (\bibinfo {year} {2017}{\natexlab{a}})},\
  \Eprint {http://arxiv.org/abs/1712.02229} {arXiv:1712.02229 [hep-ph]}
  \BibitemShut {NoStop}%
%%CITATION = ARXIV:1712.02229;%%
\bibitem [{\citenamefont {Abreu}\ \emph {et~al.}(2017)\citenamefont {Abreu},
  \citenamefont {Febres~Cordero}, \citenamefont {Ita}, \citenamefont {Page},\
  and\ \citenamefont {Zeng}}]{Abreu:2017hqn}%
  \BibitemOpen
  \bibfield  {author} {\bibinfo {author} {\bibfnamefont {S.}~\bibnamefont
  {Abreu}}, \bibinfo {author} {\bibfnamefont {F.}~\bibnamefont
  {Febres~Cordero}}, \bibinfo {author} {\bibfnamefont {H.}~\bibnamefont {Ita}},
  \bibinfo {author} {\bibfnamefont {B.}~\bibnamefont {Page}}, \ and\ \bibinfo
  {author} {\bibfnamefont {M.}~\bibnamefont {Zeng}},\ }\href@noop {} {\
  (\bibinfo {year} {2017})},\ \Eprint {http://arxiv.org/abs/1712.03946}
  {arXiv:1712.03946 [hep-ph]} \BibitemShut {NoStop}%
%%CITATION = ARXIV:1712.03946;%%
\bibitem [{\citenamefont {Badger}\ \emph {et~al.}(2013)\citenamefont {Badger},
  \citenamefont {Frellesvig},\ and\ \citenamefont {Zhang}}]{Badger:2013gxa}%
  \BibitemOpen
  \bibfield  {author} {\bibinfo {author} {\bibfnamefont {S.}~\bibnamefont
  {Badger}}, \bibinfo {author} {\bibfnamefont {H.}~\bibnamefont {Frellesvig}},
  \ and\ \bibinfo {author} {\bibfnamefont {Y.}~\bibnamefont {Zhang}},\ }\href
  {\doibase 10.1007/JHEP12(2013)045} {\bibfield  {journal} {\bibinfo  {journal}
  {JHEP}\ }\textbf {\bibinfo {volume} {12}},\ \bibinfo {pages} {045} (\bibinfo
  {year} {2013})},\ \Eprint {http://arxiv.org/abs/1310.1051} {arXiv:1310.1051
  [hep-ph]} \BibitemShut {NoStop}%
%%CITATION = ARXIV:1310.1051;%%
\bibitem [{\citenamefont {Badger}\ \emph {et~al.}(2016)\citenamefont {Badger},
  \citenamefont {Mogull},\ and\ \citenamefont {Peraro}}]{Badger:2016ozq}%
  \BibitemOpen
  \bibfield  {author} {\bibinfo {author} {\bibfnamefont {S.}~\bibnamefont
  {Badger}}, \bibinfo {author} {\bibfnamefont {G.}~\bibnamefont {Mogull}}, \
  and\ \bibinfo {author} {\bibfnamefont {T.}~\bibnamefont {Peraro}},\ }\href
  {\doibase 10.1007/JHEP08(2016)063} {\bibfield  {journal} {\bibinfo  {journal}
  {JHEP}\ }\textbf {\bibinfo {volume} {08}},\ \bibinfo {pages} {063} (\bibinfo
  {year} {2016})},\ \Eprint {http://arxiv.org/abs/1606.02244} {arXiv:1606.02244
  [hep-ph]} \BibitemShut {NoStop}%
%%CITATION = ARXIV:1606.02244;%%
\bibitem [{\citenamefont {Dunbar}\ \emph {et~al.}(2017)\citenamefont {Dunbar},
  \citenamefont {Godwin}, \citenamefont {Jehu},\ and\ \citenamefont
  {Perkins}}]{Dunbar:2017nfy}%
  \BibitemOpen
  \bibfield  {author} {\bibinfo {author} {\bibfnamefont {D.~C.}\ \bibnamefont
  {Dunbar}}, \bibinfo {author} {\bibfnamefont {J.~H.}\ \bibnamefont {Godwin}},
  \bibinfo {author} {\bibfnamefont {G.~R.}\ \bibnamefont {Jehu}}, \ and\
  \bibinfo {author} {\bibfnamefont {W.~B.}\ \bibnamefont {Perkins}},\ }\href
  {\doibase 10.1103/PhysRevD.96.116013} {\bibfield  {journal} {\bibinfo
  {journal} {Phys. Rev.}\ }\textbf {\bibinfo {volume} {D96}},\ \bibinfo {pages}
  {116013} (\bibinfo {year} {2017})},\ \Eprint
  {http://arxiv.org/abs/1710.10071} {arXiv:1710.10071 [hep-th]} \BibitemShut
  {NoStop}%
%%CITATION = ARXIV:1710.10071;%%
\bibitem [{\citenamefont {Kawai}\ \emph {et~al.}(1986)\citenamefont {Kawai},
  \citenamefont {Lewellen},\ and\ \citenamefont {Tye}}]{Kawai:1985xq}%
  \BibitemOpen
  \bibfield  {author} {\bibinfo {author} {\bibfnamefont {H.}~\bibnamefont
  {Kawai}}, \bibinfo {author} {\bibfnamefont {D.~C.}\ \bibnamefont {Lewellen}},
  \ and\ \bibinfo {author} {\bibfnamefont {S.~H.~H.}\ \bibnamefont {Tye}},\
  }\href {\doibase 10.1016/0550-3213(86)90362-7} {\bibfield  {journal}
  {\bibinfo  {journal} {Nucl. Phys.}\ }\textbf {\bibinfo {volume} {B269}},\
  \bibinfo {pages} {1} (\bibinfo {year} {1986})}\BibitemShut {NoStop}%
%%CITATION = NUPHA,B269,1;%%
\bibitem [{\citenamefont {Bern}\ \emph {et~al.}(2008)\citenamefont {Bern},
  \citenamefont {Carrasco},\ and\ \citenamefont {Johansson}}]{Bern:2008qj}%
  \BibitemOpen
  \bibfield  {author} {\bibinfo {author} {\bibfnamefont {Z.}~\bibnamefont
  {Bern}}, \bibinfo {author} {\bibfnamefont {J.~J.~M.}\ \bibnamefont
  {Carrasco}}, \ and\ \bibinfo {author} {\bibfnamefont {H.}~\bibnamefont
  {Johansson}},\ }\href {\doibase 10.1103/PhysRevD.78.085011} {\bibfield
  {journal} {\bibinfo  {journal} {Phys. Rev.}\ }\textbf {\bibinfo {volume}
  {D78}},\ \bibinfo {pages} {085011} (\bibinfo {year} {2008})},\ \Eprint
  {http://arxiv.org/abs/0805.3993} {arXiv:0805.3993 [hep-ph]} \BibitemShut
  {NoStop}%
%%CITATION = ARXIV:0805.3993;%%
\bibitem [{\citenamefont {Arkani-Hamed}\ \emph {et~al.}(2016)\citenamefont
  {Arkani-Hamed}, \citenamefont {Rodina},\ and\ \citenamefont
  {Trnka}}]{Arkani-Hamed:2016rak}%
  \BibitemOpen
  \bibfield  {author} {\bibinfo {author} {\bibfnamefont {N.}~\bibnamefont
  {Arkani-Hamed}}, \bibinfo {author} {\bibfnamefont {L.}~\bibnamefont
  {Rodina}}, \ and\ \bibinfo {author} {\bibfnamefont {J.}~\bibnamefont
  {Trnka}},\ }\href@noop {} {\  (\bibinfo {year} {2016})},\ \Eprint
  {http://arxiv.org/abs/1612.02797} {arXiv:1612.02797 [hep-th]} \BibitemShut
  {NoStop}%
%%CITATION = ARXIV:1612.02797;%%
\bibitem [{\citenamefont {Rodina}(2016)}]{Rodina:2016mbk}%
  \BibitemOpen
  \bibfield  {author} {\bibinfo {author} {\bibfnamefont {L.}~\bibnamefont
  {Rodina}},\ }\href@noop {} {\  (\bibinfo {year} {2016})},\ \Eprint
  {http://arxiv.org/abs/1612.03885} {arXiv:1612.03885 [hep-th]} \BibitemShut
  {NoStop}%
%%CITATION = ARXIV:1612.03885;%%
\bibitem [{\citenamefont {Boels}\ and\ \citenamefont
  {Medina}(2017)}]{Boels:2016xhc}%
  \BibitemOpen
  \bibfield  {author} {\bibinfo {author} {\bibfnamefont {R.~H.}\ \bibnamefont
  {Boels}}\ and\ \bibinfo {author} {\bibfnamefont {R.}~\bibnamefont {Medina}},\
  }\href {\doibase 10.1103/PhysRevLett.118.061602} {\bibfield  {journal}
  {\bibinfo  {journal} {Phys. Rev. Lett.}\ }\textbf {\bibinfo {volume} {118}},\
  \bibinfo {pages} {061602} (\bibinfo {year} {2017})},\ \Eprint
  {http://arxiv.org/abs/1607.08246} {arXiv:1607.08246 [hep-th]} \BibitemShut
  {NoStop}%
%%CITATION = ARXIV:1607.08246;%%
\bibitem [{\citenamefont {Barreiro}\ and\ \citenamefont
  {Medina}(2014)}]{Barreiro:2013dpa}%
  \BibitemOpen
  \bibfield  {author} {\bibinfo {author} {\bibfnamefont {L.~A.}\ \bibnamefont
  {Barreiro}}\ and\ \bibinfo {author} {\bibfnamefont {R.}~\bibnamefont
  {Medina}},\ }\href {\doibase 10.1016/j.nuclphysb.2014.07.015} {\bibfield
  {journal} {\bibinfo  {journal} {Nucl. Phys.}\ }\textbf {\bibinfo {volume}
  {B886}},\ \bibinfo {pages} {870} (\bibinfo {year} {2014})},\ \Eprint
  {http://arxiv.org/abs/1310.5942} {arXiv:1310.5942 [hep-th]} \BibitemShut
  {NoStop}%
%%CITATION = ARXIV:1310.5942;%%
\bibitem [{\citenamefont {Boels}\ and\ \citenamefont
  {Luo}(2017)}]{Boels:2017gyc}%
  \BibitemOpen
  \bibfield  {author} {\bibinfo {author} {\bibfnamefont {R.~H.}\ \bibnamefont
  {Boels}}\ and\ \bibinfo {author} {\bibfnamefont {H.}~\bibnamefont {Luo}},\
  }\href@noop {} {\  (\bibinfo {year} {2017})},\ \Eprint
  {http://arxiv.org/abs/1710.10208} {arXiv:1710.10208 [hep-th]} \BibitemShut
  {NoStop}%
%%CITATION = ARXIV:1710.10208;%%
\bibitem [{\citenamefont {Bern}\ \emph {et~al.}(2017)\citenamefont {Bern},
  \citenamefont {Edison}, \citenamefont {Kosower},\ and\ \citenamefont
  {Parra-Martinez}}]{Bern:2017tuc}%
  \BibitemOpen
  \bibfield  {author} {\bibinfo {author} {\bibfnamefont {Z.}~\bibnamefont
  {Bern}}, \bibinfo {author} {\bibfnamefont {A.}~\bibnamefont {Edison}},
  \bibinfo {author} {\bibfnamefont {D.}~\bibnamefont {Kosower}}, \ and\
  \bibinfo {author} {\bibfnamefont {J.}~\bibnamefont {Parra-Martinez}},\ }\href
  {\doibase 10.1103/PhysRevD.96.066004} {\bibfield  {journal} {\bibinfo
  {journal} {Phys. Rev.}\ }\textbf {\bibinfo {volume} {D96}},\ \bibinfo {pages}
  {066004} (\bibinfo {year} {2017})},\ \Eprint
  {http://arxiv.org/abs/1706.01486} {arXiv:1706.01486 [hep-th]} \BibitemShut
  {NoStop}%
%%CITATION = ARXIV:1706.01486;%%
\bibitem [{\citenamefont {Glover}\ and\ \citenamefont
  {Tejeda-Yeomans}(2003)}]{Glover:2003cm}%
  \BibitemOpen
  \bibfield  {author} {\bibinfo {author} {\bibfnamefont {E.~W.~N.}\
  \bibnamefont {Glover}}\ and\ \bibinfo {author} {\bibfnamefont {M.~E.}\
  \bibnamefont {Tejeda-Yeomans}},\ }\href {\doibase
  10.1088/1126-6708/2003/06/033} {\bibfield  {journal} {\bibinfo  {journal}
  {JHEP}\ }\textbf {\bibinfo {volume} {06}},\ \bibinfo {pages} {033} (\bibinfo
  {year} {2003})},\ \Eprint {http://arxiv.org/abs/hep-ph/0304169}
  {arXiv:hep-ph/0304169 [hep-ph]} \BibitemShut {NoStop}%
%%CITATION = HEP-PH/0304169;%%
\bibitem [{\citenamefont {Gehrmann}\ \emph {et~al.}(2012)\citenamefont
  {Gehrmann}, \citenamefont {Jaquier}, \citenamefont {Glover},\ and\
  \citenamefont {Koukoutsakis}}]{Gehrmann:2011aa}%
  \BibitemOpen
  \bibfield  {author} {\bibinfo {author} {\bibfnamefont {T.}~\bibnamefont
  {Gehrmann}}, \bibinfo {author} {\bibfnamefont {M.}~\bibnamefont {Jaquier}},
  \bibinfo {author} {\bibfnamefont {E.~W.~N.}\ \bibnamefont {Glover}}, \ and\
  \bibinfo {author} {\bibfnamefont {A.}~\bibnamefont {Koukoutsakis}},\ }\href
  {\doibase 10.1007/JHEP02(2012)056} {\bibfield  {journal} {\bibinfo  {journal}
  {JHEP}\ }\textbf {\bibinfo {volume} {02}},\ \bibinfo {pages} {056} (\bibinfo
  {year} {2012})},\ \Eprint {http://arxiv.org/abs/1112.3554} {arXiv:1112.3554
  [hep-ph]} \BibitemShut {NoStop}%
%%CITATION = ARXIV:1112.3554;%%
\bibitem [{\citenamefont {Peskin}\ and\ \citenamefont
  {Schroeder}(1995)}]{Peskin:1995ev}%
  \BibitemOpen
  \bibfield  {author} {\bibinfo {author} {\bibfnamefont {M.~E.}\ \bibnamefont
  {Peskin}}\ and\ \bibinfo {author} {\bibfnamefont {D.~V.}\ \bibnamefont
  {Schroeder}},\ }\href {http://www.slac.stanford.edu/~mpeskin/QFT.html} {\emph
  {\bibinfo {title} {{An Introduction to quantum field theory}}}}\ (\bibinfo
  {publisher} {Addison-Wesley},\ \bibinfo {address} {Reading, USA},\ \bibinfo
  {year} {1995})\BibitemShut {NoStop}%
%%CITATION = INSPIRE-407703;%%
\bibitem [{\citenamefont {Noether}(1918)}]{Noether:1918zz}%
  \BibitemOpen
  \bibfield  {author} {\bibinfo {author} {\bibfnamefont {E.}~\bibnamefont
  {Noether}},\ }\href {\doibase 10.1080/00411457108231446} {\bibfield
  {journal} {\bibinfo  {journal} {Gott. Nachr.}\ }\textbf {\bibinfo {volume}
  {1918}},\ \bibinfo {pages} {235} (\bibinfo {year} {1918})},\ \bibinfo {note}
  {[Transp. Theory Statist. Phys.1,186(1971)]},\ \Eprint
  {http://arxiv.org/abs/physics/0503066} {arXiv:physics/0503066 [physics]}
  \BibitemShut {NoStop}%
%%CITATION = PHYSICS/0503066;%%
\bibitem [{\citenamefont {Chetyrkin}\ and\ \citenamefont
  {Tkachov}(1981)}]{Chetyrkin:1981qh}%
  \BibitemOpen
  \bibfield  {author} {\bibinfo {author} {\bibfnamefont {K.~G.}\ \bibnamefont
  {Chetyrkin}}\ and\ \bibinfo {author} {\bibfnamefont {F.~V.}\ \bibnamefont
  {Tkachov}},\ }\href {\doibase 10.1016/0550-3213(81)90199-1} {\bibfield
  {journal} {\bibinfo  {journal} {Nucl. Phys.}\ }\textbf {\bibinfo {volume}
  {B192}},\ \bibinfo {pages} {159} (\bibinfo {year} {1981})}\BibitemShut
  {NoStop}%
%%CITATION = NUPHA,B192,159;%%
\bibitem [{\citenamefont {Laporta}(2000)}]{Laporta:2001dd}%
  \BibitemOpen
  \bibfield  {author} {\bibinfo {author} {\bibfnamefont {S.}~\bibnamefont
  {Laporta}},\ }\href {\doibase 10.1016/S0217-751X(00)00215-7,
  10.1142/S0217751X00002157} {\bibfield  {journal} {\bibinfo  {journal} {Int.
  J. Mod. Phys.}\ }\textbf {\bibinfo {volume} {A15}},\ \bibinfo {pages} {5087}
  (\bibinfo {year} {2000})},\ \Eprint {http://arxiv.org/abs/hep-ph/0102033}
  {arXiv:hep-ph/0102033 [hep-ph]} \BibitemShut {NoStop}%
%%CITATION = HEP-PH/0102033;%%
\bibitem [{\citenamefont {Smirnov}(2008)}]{Smirnov:2008iw}%
  \BibitemOpen
  \bibfield  {author} {\bibinfo {author} {\bibfnamefont {A.~V.}\ \bibnamefont
  {Smirnov}},\ }\href {\doibase 10.1088/1126-6708/2008/10/107} {\bibfield
  {journal} {\bibinfo  {journal} {JHEP}\ }\textbf {\bibinfo {volume} {10}},\
  \bibinfo {pages} {107} (\bibinfo {year} {2008})},\ \Eprint
  {http://arxiv.org/abs/0807.3243} {arXiv:0807.3243 [hep-ph]} \BibitemShut
  {NoStop}%
%%CITATION = ARXIV:0807.3243;%%
\bibitem [{\citenamefont {Smirnov}\ and\ \citenamefont
  {Smirnov}(2013)}]{Smirnov:2013dia}%
  \BibitemOpen
  \bibfield  {author} {\bibinfo {author} {\bibfnamefont {A.~V.}\ \bibnamefont
  {Smirnov}}\ and\ \bibinfo {author} {\bibfnamefont {V.~A.}\ \bibnamefont
  {Smirnov}},\ }\href {\doibase 10.1016/j.cpc.2013.06.016} {\bibfield
  {journal} {\bibinfo  {journal} {Comput. Phys. Commun.}\ }\textbf {\bibinfo
  {volume} {184}},\ \bibinfo {pages} {2820} (\bibinfo {year} {2013})},\ \Eprint
  {http://arxiv.org/abs/1302.5885} {arXiv:1302.5885 [hep-ph]} \BibitemShut
  {NoStop}%
%%CITATION = ARXIV:1302.5885;%%
\bibitem [{\citenamefont {Smirnov}(2015)}]{Smirnov:2014hma}%
  \BibitemOpen
  \bibfield  {author} {\bibinfo {author} {\bibfnamefont {A.~V.}\ \bibnamefont
  {Smirnov}},\ }\href {\doibase 10.1016/j.cpc.2014.11.024} {\bibfield
  {journal} {\bibinfo  {journal} {Comput. Phys. Commun.}\ }\textbf {\bibinfo
  {volume} {189}},\ \bibinfo {pages} {182} (\bibinfo {year} {2015})},\ \Eprint
  {http://arxiv.org/abs/1408.2372} {arXiv:1408.2372 [hep-ph]} \BibitemShut
  {NoStop}%
%%CITATION = ARXIV:1408.2372;%%
\bibitem [{\citenamefont {Maierhoefer}\ \emph {et~al.}(2017)\citenamefont
  {Maierhoefer}, \citenamefont {Usovitsch},\ and\ \citenamefont
  {Uwer}}]{Maierhoefer:2017hyi}%
  \BibitemOpen
  \bibfield  {author} {\bibinfo {author} {\bibfnamefont {P.}~\bibnamefont
  {Maierhoefer}}, \bibinfo {author} {\bibfnamefont {J.}~\bibnamefont
  {Usovitsch}}, \ and\ \bibinfo {author} {\bibfnamefont {P.}~\bibnamefont
  {Uwer}},\ }\href@noop {} {\  (\bibinfo {year} {2017})},\ \Eprint
  {http://arxiv.org/abs/1705.05610} {arXiv:1705.05610 [hep-ph]} \BibitemShut
  {NoStop}%
%%CITATION = ARXIV:1705.05610;%%
\bibitem [{\citenamefont {von Manteuffel}\ and\ \citenamefont
  {Studerus}(2012)}]{vonManteuffel:2012np}%
  \BibitemOpen
  \bibfield  {author} {\bibinfo {author} {\bibfnamefont {A.}~\bibnamefont {von
  Manteuffel}}\ and\ \bibinfo {author} {\bibfnamefont {C.}~\bibnamefont
  {Studerus}},\ }\href@noop {} {\  (\bibinfo {year} {2012})},\ \Eprint
  {http://arxiv.org/abs/1201.4330} {arXiv:1201.4330 [hep-ph]} \BibitemShut
  {NoStop}%
%%CITATION = ARXIV:1201.4330;%%
\bibitem [{\citenamefont {Lee}(2012)}]{Lee:2012cn}%
  \BibitemOpen
  \bibfield  {author} {\bibinfo {author} {\bibfnamefont {R.~N.}\ \bibnamefont
  {Lee}},\ }\href@noop {} {\  (\bibinfo {year} {2012})},\ \Eprint
  {http://arxiv.org/abs/1212.2685} {arXiv:1212.2685 [hep-ph]} \BibitemShut
  {NoStop}%
%%CITATION = ARXIV:1212.2685;%%
\bibitem [{\citenamefont {Lee}(2014)}]{Lee:2013mka}%
  \BibitemOpen
  \bibfield  {author} {\bibinfo {author} {\bibfnamefont {R.~N.}\ \bibnamefont
  {Lee}},\ }\bibfield  {booktitle} {\emph {\bibinfo {booktitle} {{Proceedings,
  15th International Workshop on Advanced Computing and Analysis Techniques in
  Physics Research (ACAT 2013): Beijing, China, May 16-21, 2013}}},\ }\href
  {\doibase 10.1088/1742-6596/523/1/012059} {\bibfield  {journal} {\bibinfo
  {journal} {J. Phys. Conf. Ser.}\ }\textbf {\bibinfo {volume} {523}},\
  \bibinfo {pages} {012059} (\bibinfo {year} {2014})},\ \Eprint
  {http://arxiv.org/abs/1310.1145} {arXiv:1310.1145 [hep-ph]} \BibitemShut
  {NoStop}%
%%CITATION = ARXIV:1310.1145;%%
\bibitem [{\citenamefont {Bjerrum-Bohr}\ \emph {et~al.}(2008)\citenamefont
  {Bjerrum-Bohr}, \citenamefont {Dunbar},\ and\ \citenamefont
  {Perkins}}]{BjerrumBohr:2007vu}%
  \BibitemOpen
  \bibfield  {author} {\bibinfo {author} {\bibfnamefont {N.~E.~J.}\
  \bibnamefont {Bjerrum-Bohr}}, \bibinfo {author} {\bibfnamefont {D.~C.}\
  \bibnamefont {Dunbar}}, \ and\ \bibinfo {author} {\bibfnamefont {W.~B.}\
  \bibnamefont {Perkins}},\ }\href {\doibase 10.1088/1126-6708/2008/04/038}
  {\bibfield  {journal} {\bibinfo  {journal} {JHEP}\ }\textbf {\bibinfo
  {volume} {04}},\ \bibinfo {pages} {038} (\bibinfo {year} {2008})},\ \Eprint
  {http://arxiv.org/abs/0709.2086} {arXiv:0709.2086 [hep-ph]} \BibitemShut
  {NoStop}%
%%CITATION = ARXIV:0709.2086;%%
\bibitem [{\citenamefont {Henn}(2015)}]{Henn:2014qga}%
  \BibitemOpen
  \bibfield  {author} {\bibinfo {author} {\bibfnamefont {J.~M.}\ \bibnamefont
  {Henn}},\ }\href {\doibase 10.1088/1751-8113/48/15/153001} {\bibfield
  {journal} {\bibinfo  {journal} {J. Phys.}\ }\textbf {\bibinfo {volume}
  {A48}},\ \bibinfo {pages} {153001} (\bibinfo {year} {2015})},\ \Eprint
  {http://arxiv.org/abs/1412.2296} {arXiv:1412.2296 [hep-ph]} \BibitemShut
  {NoStop}%
%%CITATION = ARXIV:1412.2296;%%
\bibitem [{\citenamefont {Badger}\ \emph
  {et~al.}(2017{\natexlab{b}})\citenamefont {Badger}, \citenamefont
  {Br{\o}nnum-Hansen}, \citenamefont {Buciuni},\ and\ \citenamefont
  {O'Connell}}]{Badger:2017gta}%
  \BibitemOpen
  \bibfield  {author} {\bibinfo {author} {\bibfnamefont {S.}~\bibnamefont
  {Badger}}, \bibinfo {author} {\bibfnamefont {C.}~\bibnamefont
  {Br{\o}nnum-Hansen}}, \bibinfo {author} {\bibfnamefont {F.}~\bibnamefont
  {Buciuni}}, \ and\ \bibinfo {author} {\bibfnamefont {D.}~\bibnamefont
  {O'Connell}},\ }\href {\doibase 10.1007/JHEP06(2017)141} {\bibfield
  {journal} {\bibinfo  {journal} {JHEP}\ }\textbf {\bibinfo {volume} {06}},\
  \bibinfo {pages} {141} (\bibinfo {year} {2017}{\natexlab{b}})},\ \Eprint
  {http://arxiv.org/abs/1703.05734} {arXiv:1703.05734 [hep-ph]} \BibitemShut
  {NoStop}%
%%CITATION = ARXIV:1703.05734;%%
\bibitem [{\citenamefont {Henn}(2013)}]{Henn:2013pwa}%
  \BibitemOpen
  \bibfield  {author} {\bibinfo {author} {\bibfnamefont {J.~M.}\ \bibnamefont
  {Henn}},\ }\href {\doibase 10.1103/PhysRevLett.110.251601} {\bibfield
  {journal} {\bibinfo  {journal} {Phys. Rev. Lett.}\ }\textbf {\bibinfo
  {volume} {110}},\ \bibinfo {pages} {251601} (\bibinfo {year} {2013})},\
  \Eprint {http://arxiv.org/abs/1304.1806} {arXiv:1304.1806 [hep-th]}
  \BibitemShut {NoStop}%
%%CITATION = ARXIV:1304.1806;%%
\bibitem [{\citenamefont {Gituliar}\ and\ \citenamefont
  {Magerya}(2017)}]{Gituliar:2017vzm}%
  \BibitemOpen
  \bibfield  {author} {\bibinfo {author} {\bibfnamefont {O.}~\bibnamefont
  {Gituliar}}\ and\ \bibinfo {author} {\bibfnamefont {V.}~\bibnamefont
  {Magerya}},\ }\href {\doibase 10.1016/j.cpc.2017.05.004} {\bibfield
  {journal} {\bibinfo  {journal} {Comput. Phys. Commun.}\ }\textbf {\bibinfo
  {volume} {219}},\ \bibinfo {pages} {329} (\bibinfo {year} {2017})},\ \Eprint
  {http://arxiv.org/abs/1701.04269} {arXiv:1701.04269 [hep-ph]} \BibitemShut
  {NoStop}%
%%CITATION = ARXIV:1701.04269;%%
\bibitem [{\citenamefont {Bern}\ \emph {et~al.}(1995)\citenamefont {Bern},
  \citenamefont {Dixon}, \citenamefont {Dunbar},\ and\ \citenamefont
  {Kosower}}]{Bern:1994cg}%
  \BibitemOpen
  \bibfield  {author} {\bibinfo {author} {\bibfnamefont {Z.}~\bibnamefont
  {Bern}}, \bibinfo {author} {\bibfnamefont {L.~J.}\ \bibnamefont {Dixon}},
  \bibinfo {author} {\bibfnamefont {D.~C.}\ \bibnamefont {Dunbar}}, \ and\
  \bibinfo {author} {\bibfnamefont {D.~A.}\ \bibnamefont {Kosower}},\ }\href
  {\doibase 10.1016/0550-3213(94)00488-Z} {\bibfield  {journal} {\bibinfo
  {journal} {Nucl. Phys.}\ }\textbf {\bibinfo {volume} {B435}},\ \bibinfo
  {pages} {59} (\bibinfo {year} {1995})},\ \Eprint
  {http://arxiv.org/abs/hep-ph/9409265} {arXiv:hep-ph/9409265 [hep-ph]}
  \BibitemShut {NoStop}%
%%CITATION = HEP-PH/9409265;%%
\bibitem [{\citenamefont {Gehrmann}\ \emph {et~al.}(2016)\citenamefont
  {Gehrmann}, \citenamefont {Henn},\ and\ \citenamefont
  {Lo~Presti}}]{Gehrmann:2015bfy}%
  \BibitemOpen
  \bibfield  {author} {\bibinfo {author} {\bibfnamefont {T.}~\bibnamefont
  {Gehrmann}}, \bibinfo {author} {\bibfnamefont {J.~M.}\ \bibnamefont {Henn}},
  \ and\ \bibinfo {author} {\bibfnamefont {N.~A.}\ \bibnamefont {Lo~Presti}},\
  }\href {\doibase 10.1103/PhysRevLett.116.189903,
  10.1103/PhysRevLett.116.062001} {\bibfield  {journal} {\bibinfo  {journal}
  {Phys. Rev. Lett.}\ }\textbf {\bibinfo {volume} {116}},\ \bibinfo {pages}
  {062001} (\bibinfo {year} {2016})},\ \bibinfo {note} {[Erratum: Phys. Rev.
  Lett.116,no.18,189903(2016)]},\ \Eprint {http://arxiv.org/abs/1511.05409}
  {arXiv:1511.05409 [hep-ph]} \BibitemShut {NoStop}%
%%CITATION = ARXIV:1511.05409;%%
\bibitem [{\citenamefont {Papadopoulos}\ \emph {et~al.}(2016)\citenamefont
  {Papadopoulos}, \citenamefont {Tommasini},\ and\ \citenamefont
  {Wever}}]{Papadopoulos:2015jft}%
  \BibitemOpen
  \bibfield  {author} {\bibinfo {author} {\bibfnamefont {C.~G.}\ \bibnamefont
  {Papadopoulos}}, \bibinfo {author} {\bibfnamefont {D.}~\bibnamefont
  {Tommasini}}, \ and\ \bibinfo {author} {\bibfnamefont {C.}~\bibnamefont
  {Wever}},\ }\href {\doibase 10.1007/JHEP04(2016)078} {\bibfield  {journal}
  {\bibinfo  {journal} {JHEP}\ }\textbf {\bibinfo {volume} {04}},\ \bibinfo
  {pages} {078} (\bibinfo {year} {2016})},\ \Eprint
  {http://arxiv.org/abs/1511.09404} {arXiv:1511.09404 [hep-ph]} \BibitemShut
  {NoStop}%
%%CITATION = ARXIV:1511.09404;%%
\bibitem [{\citenamefont {Henn}\ \emph {et~al.}(2013)\citenamefont {Henn},
  \citenamefont {Smirnov},\ and\ \citenamefont {Smirnov}}]{Henn:2013fah}%
  \BibitemOpen
  \bibfield  {author} {\bibinfo {author} {\bibfnamefont {J.~M.}\ \bibnamefont
  {Henn}}, \bibinfo {author} {\bibfnamefont {A.~V.}\ \bibnamefont {Smirnov}}, \
  and\ \bibinfo {author} {\bibfnamefont {V.~A.}\ \bibnamefont {Smirnov}},\
  }\href {\doibase 10.1007/JHEP07(2013)128} {\bibfield  {journal} {\bibinfo
  {journal} {JHEP}\ }\textbf {\bibinfo {volume} {07}},\ \bibinfo {pages} {128}
  (\bibinfo {year} {2013})},\ \Eprint {http://arxiv.org/abs/1306.2799}
  {arXiv:1306.2799 [hep-th]} \BibitemShut {NoStop}%
%%CITATION = ARXIV:1306.2799;%%
\bibitem [{\citenamefont {Arkani-Hamed}\ and\ \citenamefont
  {Trnka}(2014)}]{Arkani-Hamed:2013jha}%
  \BibitemOpen
  \bibfield  {author} {\bibinfo {author} {\bibfnamefont {N.}~\bibnamefont
  {Arkani-Hamed}}\ and\ \bibinfo {author} {\bibfnamefont {J.}~\bibnamefont
  {Trnka}},\ }\href {\doibase 10.1007/JHEP10(2014)030} {\bibfield  {journal}
  {\bibinfo  {journal} {JHEP}\ }\textbf {\bibinfo {volume} {10}},\ \bibinfo
  {pages} {030} (\bibinfo {year} {2014})},\ \Eprint
  {http://arxiv.org/abs/1312.2007} {arXiv:1312.2007 [hep-th]} \BibitemShut
  {NoStop}%
%%CITATION = ARXIV:1312.2007;%%
\bibitem [{\citenamefont {Cachazo}(2008)}]{Cachazo:2008vp}%
  \BibitemOpen
  \bibfield  {author} {\bibinfo {author} {\bibfnamefont {F.}~\bibnamefont
  {Cachazo}},\ }\href@noop {} {\  (\bibinfo {year} {2008})},\ \Eprint
  {http://arxiv.org/abs/0803.1988} {arXiv:0803.1988 [hep-th]} \BibitemShut
  {NoStop}%
%%CITATION = ARXIV:0803.1988;%%
\bibitem [{\citenamefont {Czakon}(2005)}]{diagen}%
  \BibitemOpen
  \bibfield  {author} {\bibinfo {author} {\bibfnamefont {M.}~\bibnamefont
  {Czakon}},\ }\href@noop {} {\enquote {\bibinfo {title} {{D}ia{G}en},}\
  }\bibinfo {howpublished}
  {{http://www-zeuthen.desy.de/theory/capp2005/Course/czakon/}} (\bibinfo
  {year} {2005})\BibitemShut {NoStop}%
\bibitem [{\citenamefont {Ita}(2016)}]{Ita:2015tya}%
  \BibitemOpen
  \bibfield  {author} {\bibinfo {author} {\bibfnamefont {H.}~\bibnamefont
  {Ita}},\ }\href {\doibase 10.1103/PhysRevD.94.116015} {\bibfield  {journal}
  {\bibinfo  {journal} {Phys. Rev.}\ }\textbf {\bibinfo {volume} {D94}},\
  \bibinfo {pages} {116015} (\bibinfo {year} {2016})},\ \Eprint
  {http://arxiv.org/abs/1510.05626} {arXiv:1510.05626 [hep-th]} \BibitemShut
  {NoStop}%
%%CITATION = ARXIV:1510.05626;%%
\bibitem [{\citenamefont {Larsen}\ and\ \citenamefont
  {Zhang}(2016)}]{Larsen:2015ped}%
  \BibitemOpen
  \bibfield  {author} {\bibinfo {author} {\bibfnamefont {K.~J.}\ \bibnamefont
  {Larsen}}\ and\ \bibinfo {author} {\bibfnamefont {Y.}~\bibnamefont {Zhang}},\
  }\href {\doibase 10.1103/PhysRevD.93.041701} {\bibfield  {journal} {\bibinfo
  {journal} {Phys. Rev.}\ }\textbf {\bibinfo {volume} {D93}},\ \bibinfo {pages}
  {041701} (\bibinfo {year} {2016})},\ \Eprint
  {http://arxiv.org/abs/1511.01071} {arXiv:1511.01071 [hep-th]} \BibitemShut
  {NoStop}%
%%CITATION = ARXIV:1511.01071;%%
\bibitem [{\citenamefont {Georgoudis}\ \emph {et~al.}(2017)\citenamefont
  {Georgoudis}, \citenamefont {Larsen},\ and\ \citenamefont
  {Zhang}}]{Georgoudis:2016wff}%
  \BibitemOpen
  \bibfield  {author} {\bibinfo {author} {\bibfnamefont {A.}~\bibnamefont
  {Georgoudis}}, \bibinfo {author} {\bibfnamefont {K.~J.}\ \bibnamefont
  {Larsen}}, \ and\ \bibinfo {author} {\bibfnamefont {Y.}~\bibnamefont
  {Zhang}},\ }\href {\doibase 10.1016/j.cpc.2017.08.013} {\bibfield  {journal}
  {\bibinfo  {journal} {Comput. Phys. Commun.}\ }\textbf {\bibinfo {volume}
  {221}},\ \bibinfo {pages} {203} (\bibinfo {year} {2017})},\ \Eprint
  {http://arxiv.org/abs/1612.04252} {arXiv:1612.04252 [hep-th]} \BibitemShut
  {NoStop}%
%%CITATION = ARXIV:1612.04252;%%
\bibitem [{\citenamefont {Boehm}\ \emph {et~al.}(2017)\citenamefont {Boehm},
  \citenamefont {Georgoudis}, \citenamefont {Larsen}, \citenamefont {Schulze},\
  and\ \citenamefont {Zhang}}]{Boehm:2017wjc}%
  \BibitemOpen
  \bibfield  {author} {\bibinfo {author} {\bibfnamefont {J.}~\bibnamefont
  {Boehm}}, \bibinfo {author} {\bibfnamefont {A.}~\bibnamefont {Georgoudis}},
  \bibinfo {author} {\bibfnamefont {K.~J.}\ \bibnamefont {Larsen}}, \bibinfo
  {author} {\bibfnamefont {M.}~\bibnamefont {Schulze}}, \ and\ \bibinfo
  {author} {\bibfnamefont {Y.}~\bibnamefont {Zhang}},\ }\href@noop {} {\
  (\bibinfo {year} {2017})},\ \Eprint {http://arxiv.org/abs/1712.09737}
  {arXiv:1712.09737 [hep-th]} \BibitemShut {NoStop}%
%%CITATION = ARXIV:1712.09737;%%
\bibitem [{\citenamefont {Boehm}\ \emph {et~al.}(2018)\citenamefont {Boehm},
  \citenamefont {Georgoudis}, \citenamefont {Larsen}, \citenamefont
  {Schoenemann},\ and\ \citenamefont {Zhang}}]{Boehm:2018fpv}%
  \BibitemOpen
  \bibfield  {author} {\bibinfo {author} {\bibfnamefont {J.}~\bibnamefont
  {Boehm}}, \bibinfo {author} {\bibfnamefont {A.}~\bibnamefont {Georgoudis}},
  \bibinfo {author} {\bibfnamefont {K.~J.}\ \bibnamefont {Larsen}}, \bibinfo
  {author} {\bibfnamefont {H.}~\bibnamefont {Schoenemann}}, \ and\ \bibinfo
  {author} {\bibfnamefont {Y.}~\bibnamefont {Zhang}},\ }\href@noop {} {\
  (\bibinfo {year} {2018})},\ \Eprint {http://arxiv.org/abs/1805.01873}
  {arXiv:1805.01873 [hep-th]} \BibitemShut {NoStop}%
%%CITATION = ARXIV:1805.01873;%%
\end{thebibliography}%
\end{document}